\documentstyle[aps,preprint,floats,psfig]{revtex}

\draft
\begin{document}
\newcommand {\be}{\begin{equation}}
\newcommand {\ee}{\end{equation}}
\newcommand {\bea}{\begin{eqnarray}}
\newcommand {\eea}{\end{eqnarray}}
\newcommand {\nn}{\nonumber}


\title{Impurity Bound States and Symmetry of the Superconducting
Order Parameter in $\rm Sr_2RuO_4$}

\author{Kazumi Maki and Stephan Haas}
\address{Department of Physics and Astronomy, University of Southern
California, Los Angeles, CA 90089-0484
}

\date{\today}
\maketitle

\begin{abstract}

Recent experiments on $\rm Sr_2RuO_4$
have indicated the presence of linear nodes in the 
superconducting order parameter. Among the possible spin
triplet states, 2D p-wave superconductivity with
$E_u$ symmetry appears to
be inconsistent with the experiments, whereas the 2D 
f-wave order parameter with $\rm B_{1g} \times E_u$
symmetry turns out to be a more likely
candidate. Here it is shown how the quasi-particle 
bound state wave functions around a single impurity
provide a clear signature of the symmetry of the underlying
superconducting order parameter.   

\end{abstract}
\pacs{}

\section{Introduction}

In recent years, we have experienced
a rapid improvement in the quality of $\rm Sr_2RuO_4$
single crystals. This development has lead to a renewed discussion
regarding the symmetry of the underlying superconductiving order
parameter.
The data from specific heat measurements\cite{nishizaki},
NMR results on $\rm T_1^{-1}$\cite{ishida}, and magnetic
penetration depth experiments\cite{bonalde} in very pure
single crystals with $\rm T_c \simeq 1.5 K$ suggest a 
linear nodal structure in the superconducting order 
parameter, similar to the $\rm d_{x^2 - y^2}$-wave 
superconductors\cite{won}. 
Hence, the initially proposed fully gapped 2D p-wave 
superconductivity with $E_u$ symmetry\cite{rice} 
appears to be inconsistent with the experimentally observed nodal 
structure. On the other hand, the spin triplet nature 
of the superconductivity in $\rm Sr_2RuO_4$ has clearly been
established by muon rotation experiments\cite{luke} which
probe a spontaneous spin polarization, and by the flat
Knight shift seen in NMR\cite{ishida2}. 

Motivated by these recent experiments, several spin triplet
order parameters with linear nodes have been proposed.
\cite{hasegawa} In particular, from a study of the thermal conductivity 
tensor in a planar magnetic field it has been suggested that
the 2D f-wave order parameter with $\rm B_{1g} \times E_u$
symmetry is a likely candidate.\cite{dahm} Furthermore, recent
measurements of the angular dependence of the upper
critical field have detected a 
in-plane anisotropy consistent with a $\rm B_{1g} \times E_u$
component to the order parameter.\cite{mao} 

The objective of this paper is to calculate the quasi-particle
bound state wave function around non-magnetic impurities in
unconventional superconductors with the anisotropic order 
parameters which have been
proposed for $\rm Sr_2RuO_4$. It has recently been 
observed in a series of STM imaging experiments
that a Zn-impurity in the high-$\rm T_c$ compound
Bi2212 gives rise to such a bound state, leading to a distinct  
four-fold symmetric pattern in the local tunneling conductance
around the impurity.\cite{pan} 
From the theoretical side it was shown that this type of bound 
state wave function can be interpreted in terms of the solutions 
of the Bogoliubov-de Gennes equations for $\rm d_{x^2 - y^2}$-wave
superconductors.\cite{morita,haas}. Here we will perform an 
analogous analysis for the proposed 2D p-wave and f-wave 
order parameters.

\section{Bound State Wave Functions}
\subsection{2D p-wave superconductors}

For the initially proposed fully gapped odd-parity
2D p-wave superconductivity with an order parameter
$\vec{\Delta} ({\bf k}) = \Delta  \hat{z} \exp{(\pm i \phi )}$
the Bogoliubov - de Gennes equations have already been worked 
out.\cite{matsumoto,okuno} In particular, when $\rm \Delta({\bf r})
= \Delta = const. $\cite{footnote}, one obtains
\bea
E u({\bf r}) &=& \left( -\frac{\nabla^2}{2 m} - \mu - V ({\bf r}) \right)
u({\bf r}) + p_F^{-1} \Delta (i\partial_x - \partial_y )
v({\bf r}),\\
E v({\bf r}) &=& - \left( -\frac{\nabla^2}{2 m} - \mu - V ({\bf r}) \right)
v({\bf r}) + p_F^{-1} \Delta (i\partial_x + \partial_y )
u({\bf r}),
\eea
where $\mu$ is the chemical potential, and the 
impurity potential, centered at the site ${\bf r} = 0$,
is approximated by $V({\bf r}) = a \delta^2({\bf r})$.
For simplicity, only the 2D system is considered here.

A simple variational solution of Eqs. (1) and (2) is found to be 
\bea
u({\bf r}) &=&  A \exp{(- \gamma r)} J_0(p_F r), \\
v({\bf r}) &=&  A \alpha \exp{(- \gamma r)} J_1(p_F r) \exp{(i \phi)},
\eea
where $p_F$ is the Fermi momentum, $J_l(z)$ are Bessel functions
of the first kind, and $A$ is a global normalization factor. By 
inserting these variational wave functions $u({\bf r})$ and 
$v({\bf r})$ into the Bogoliubov - de Gennes equations, it follows 
that $E = -(V/2) \pm \sqrt{(K - V )^2 + \Delta^2}$ and 
$\alpha = \Delta/(E + K) \simeq 1$,
where the kinetic $K$ and the potential 
$V$ contributions to the energy are defined by
\bea
K &\equiv &  \frac{\int_0^{\infty} dr r \left[
\left( \partial_r \exp{(-\gamma r)}
J_l(p_F r)\right)^2 + \left( l \exp{(-\gamma r)} J_l(p_F r)/r \right)^2
\right]}
{2m \int_0^{\infty} dr r \left( \exp{(-\gamma r)} J_l(p_F r)\right)^2}
-\mu
\simeq \frac{\gamma^3}{m p_F},\\
V &\equiv & \frac{\int_0^{\infty} dr r \exp{(-2 \gamma r)} J_0^2(p_F r)
V({\bf r})}
{\int_0^{\infty} dr r \exp{(-2 \gamma r)} J_0^2(p_F r)}
\simeq (2 \pi \gamma p_F) \int_0^{\infty} dr r \exp{(-2 \gamma r)}
J_0^2(p_F r) V ({\bf r}).
\eea

The squares of the wave functions $u({\bf r})$ and $v({\bf r})$ can
be observed by scanning tunneling microscopy.\cite{pan}
The tunneling current,
$I({\bf r},{\it V}) \propto \int dE A_S({\bf r}, E)
A_N({\bf r}, E + e{\it	V})$, is
a convolution of the local one-particle spectral function
of the normal-state tip, $A_N ({\bf r}, E) = \sum_k
\delta (E - E_k)$, and that of the
superconducting sample, $A_S ({\bf r}, E) = \sum_k
[ |u({\bf r})|^2 \delta (E - E_k) +
  |v({\bf r})|^2 \delta (E + E_k) ]$.
The differential tunneling conductance is thus
obtained by taking the partial derivative of $I({\bf r},{\it V})$
with respect to the applied voltage ${\it V}$,
\bea
\frac{\partial I}{\partial {\it V}} ({\bf r}, {\it V}) \propto
{\rm sech}^2\left( \frac{e{\it V} - E_0}{2T} \right) |u({\bf r})|^2
+ {\rm sech}^2\left( \frac{e{\it V} + E_0}{2T} \right) |v({\bf r})|^2.
\eea
At small temperatures, the local tunneling conductance around
the impurity site is dominated by $|u({\bf r})|^2$ for a fixed binding energy
$E_0$ and by $|v({\bf r})|^2$ for $-E_0$.

In Fig. 1(a) and (b), we show $|u({\bf r})|^2$ and
$|v({\bf r})|^2$ respectively. The Fermi wave vector
$p_F  \simeq 2.7 /a$ was chosen to be consistent with
band structure calculations.\cite{mazin} The patterns 
described by the squares of these wave functions are 
concentric circles without a trace of four-fold symmetry.
If the underlying superconducting order parameter is
indeed 
$\vec{\Delta} ({\bf k}) = \Delta  \hat{z} \exp{(\pm i \phi )}$
the patterns of the local tunneling current around non-magnetic
impurities should thus be featureless along the azimuthal direction 
in the 2D plane.

\subsection{2D f-wave superconductors}

Let us now consider 2D f-wave superconductors with an
order parameter
$\vec{\Delta} ({\bf k}) = \Delta  \hat{z} \cos{(2 \phi)} \exp{(\pm i \phi )}$.
Following the above procedure, the corresponding Bogoliubov - de Gennes 
equations are then given by
\bea
E u({\bf r}) &=& \left( -\frac{\nabla^2}{2 m} - \mu - V ({\bf r}) \right)
u({\bf r}) + p_F^{-3} \Delta (\partial_x^2 - \partial_y^2 )
 (i\partial_x - \partial_y )v({\bf r}),\\
E v({\bf r}) &=& - \left( -\frac{\nabla^2}{2 m} - \mu - V ({\bf r}) \right)
v({\bf r}) + p_F^{-3} \Delta (\partial_x^2 - \partial_y^2 )
 (i\partial_x + \partial_y )u({\bf r}).
\eea

Bound state solutions for this case are found to be of the form
\bea
u({\bf r}) &=&  A \exp{(- \gamma r)}\left( J_0(p_F r) + \beta
J_4(p_F r)\cos{(4 \phi )} \right), \\
v({\bf r}) &=&  A \exp{(- \gamma r)}\left(\alpha
J_1(p_F r) \exp{(- i \phi)} + \delta J_3(p_F r) \exp{(3 i \phi)} \right).
\eea

In analogy to Ref. \cite{haas}, we obtain
\bea
E &=& K - V -\frac{1}{\sqrt{2}}\Delta ( \alpha + \delta ),  \\
E \alpha &=& - K \alpha - \frac{1}{\sqrt{2}}\Delta(1+\frac{\beta}{\sqrt{2}}),\\
E \beta &=& K \beta - \frac{1}{2} \Delta (\alpha + \delta ),  \\
E \delta &=& -K \delta - \frac{1}{\sqrt{2}}\Delta(1+\frac{\beta}{\sqrt{2}}).
\eea
From Eqs. (12) - (15) we get
$\alpha = \delta = (V - K ) / ( \sqrt{2} \Delta ) $ and
$\beta = (1 - V/(K-E))/\sqrt{2}$.
For an impurity with a scattering strength in the unitary limit
the energy of the bound state is expected to be very small,
$E \simeq 0$, and $V \simeq \Delta$. This gives $\beta \simeq 
-1/\sqrt{2}$. On the other hand, in the Born limit   
$E\simeq \Delta$ and $V \simeq 0$, which gives 
$\beta \simeq 
1/\sqrt{2}$.  
  In Fig. 2(a) and (b), we plot 
$|u({\bf r})|^2$ and $|v({\bf r})|^2$ for $\beta = 1/\sqrt{2}$,
corresponding to weak scattering.
In the limit of strong impurity scattering, $\beta = - 1/\sqrt{2}$,
the patterns are 
changed as shown in Fig. 3(a) and (b). 
It is observed that for 2D f-wave superconductors with
$\vec{\Delta} ({\bf k}) = \Delta  \hat{z} \cos{(2 \phi)} \exp{(\pm i \phi )}$,
both $|u({\bf r})|^2$ and $|v({\bf r})|^2$ have a
four-fold symmetry. Depending on the impurity scattering strength,
they extend either
in the directions of the Ru-O bonds or are tilted by an angle of 
45 degrees.
The STM imaging of impurity bound states can thus provide a 
clear signature of the order parameter symmetry for the 
underlying superconductivity. 

\section{Concluding Remarks}

Stimulated by the successful scanning tunneling microscopy imaging of
quasi-particle bound state wave functions around Zn-impurities
in Bi2212\cite{pan,hudson}, we have studied the analogous 
patterns of impurity bound states
in $\rm Sr_2RuO_4$. By applying the appropriate Bogoliubov - de Gennes
equations, the characteristic patterns were distinguished
for two proposed order parameters: (i) gapped
2D p-wave (or $E_u$) superconductors
with $\vec{\Delta} ({\bf k}) = \Delta  \hat{z} \exp{(\pm i \phi )}$,
and (ii) gapless 2D f-wave (or $\rm B_{1g} \times E_u$) superconductors 
with $\vec{\Delta} ({\bf k}) = \Delta  \hat{z} \cos{(2\phi)}
\exp{(\pm i \phi )}$. While the tunneling conductance patterns of
the fully gapped odd-parity 2D p-wave superconductor are featureless
along the azimuthal direction, a clear four-fold symmetry in
$|u({\bf r})|^2$ and $|v({\bf r})|^2$ is predicted
for the 
2D f-wave superconductor.

The experimentally observed in-plane anisotropy of the upper critical 
field $H_{c2}$ in $\rm Sr_2RuO_4$
is quite small ($\approx 3 \%$). This may
indicate that there is 
a combination of 2D f-wave and p-wave superconductivity in this
compound.\cite{mao} 
Furthermore, a possible
alternative to the plain 2D p-wave superconductivity considered
above would be a 3D $\rm A_{1g} \times E_u$ f-wave order parameter of the form 
$\vec{\Delta} ({\bf k}) = \Delta  \hat{z} \cos{(c k_z)}
\exp{(\pm i \phi )}$. The in-plane impurity bound state patterns for
2D $\rm E_u$ and 3D $\rm A_{1g} \times E_u$ 
are the same, and a directional probe along the $\hat{k}_z$-direction would 
be needed to distinguish between these two cases. 
 
Our study suggests that the Bogoliubov - de Gennes formalism in the 
continuum limit is very useful in addressing the shape of 
impurity induced bound states. The corresponding bound state 
wave functions $u({\bf r})$ and $v({\bf r})$ clearly reflect the 
symmetry of $\rm \Delta({\bf r}) $.
Therefore the imaging of these wave functions provides unique insight
into the underlying symmetry of the order parameter. A similar study
of the vortex state is in progress.

We thank Kazushige Machida,
Manfred Sigrist, and Hyekyung Won for useful discussions.
K. M. acknowledges the hospitality and the support 
of the Max-Planck Institut f\"ur Physik komplexer Systeme
at Dresden, and S. H. thanks the
Zumberge foundation for financial support.

\newpage

\begin{figure}[h]
\centerline{\psfig{figure=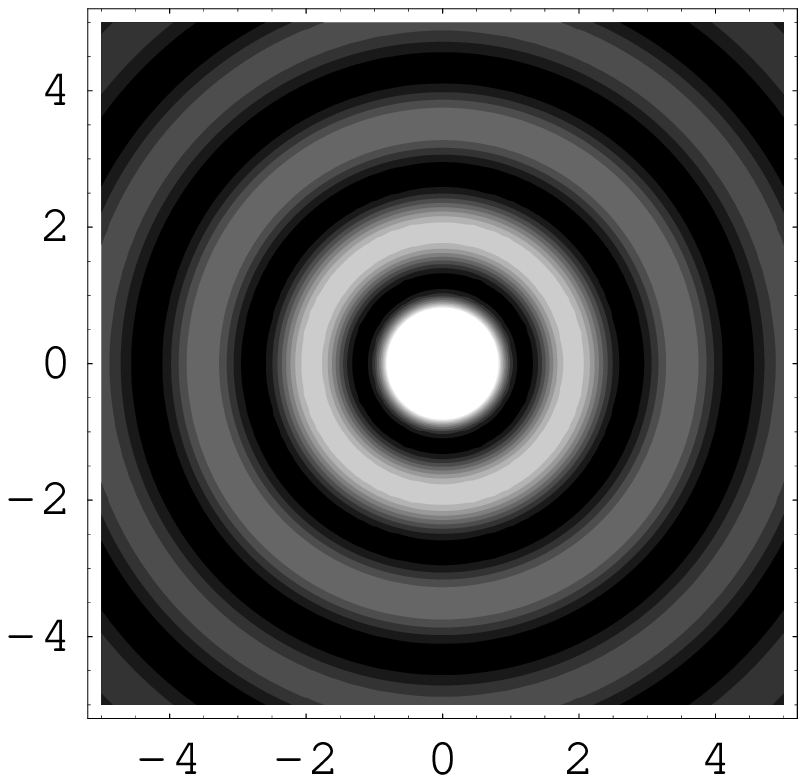,height=6cm}
\psfig{figure=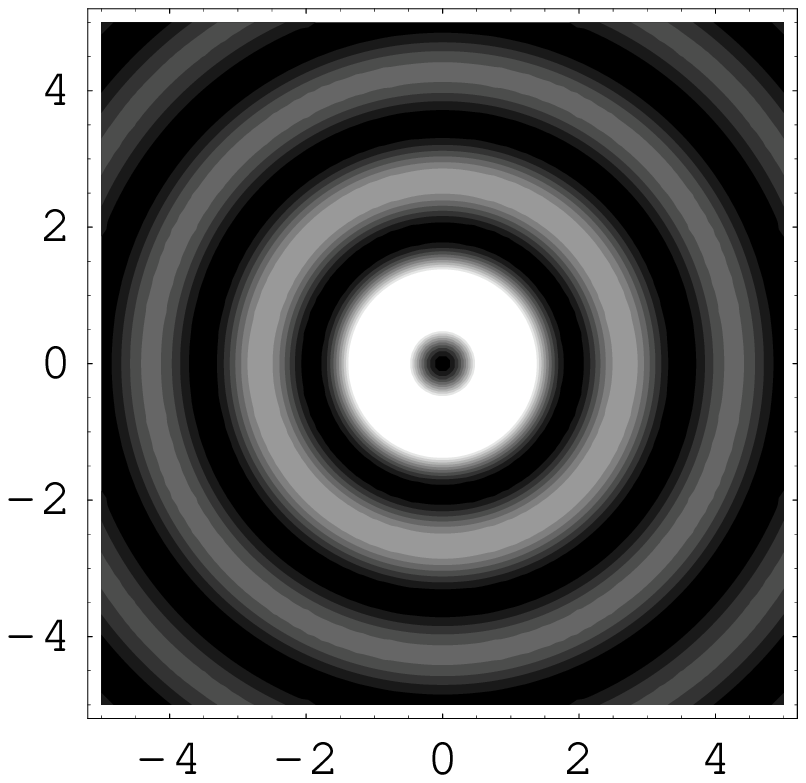,height=6cm}}
\caption{
Spatial variation of the local tunneling conductance, centered at a
non-magnetic impurity in a 2D p-wave superconductor
with $\vec{\Delta} ({\bf k}) = \Delta  \hat{z} \exp{(\pm i \phi )}$.
In the left figure, the dominant contribution
$|u({\bf r})|^2$ at the
positive bound state resonant frequency $E_0$ is shown.
On the right hand side, the dominant contribution
$|v({\bf r})|^2$ at the
negative bound state resonant frequency $-E_0$ is shown.
}
\end{figure}

\begin{figure}[h]
\centerline{\psfig{figure=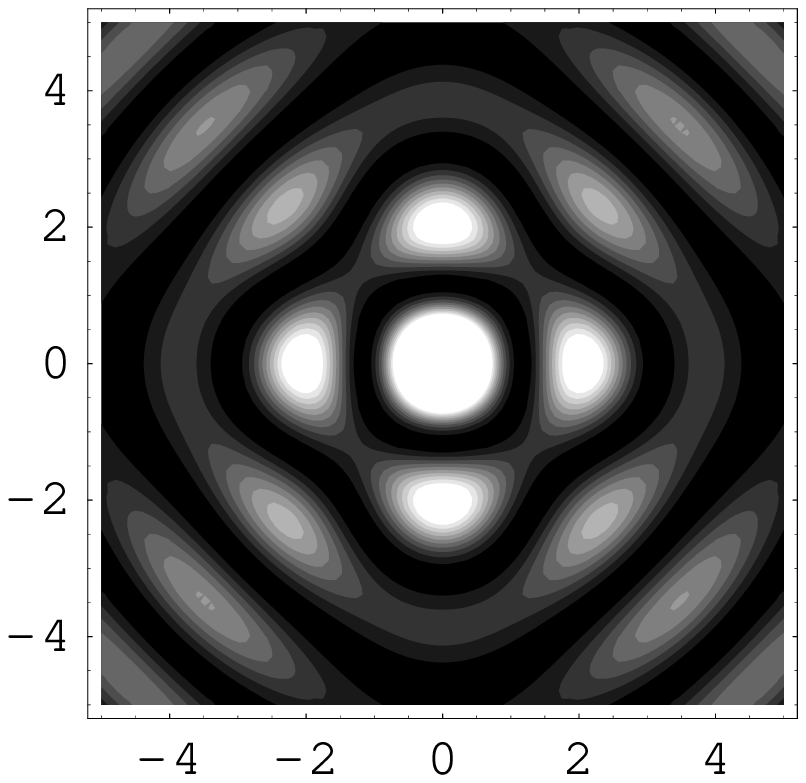,height=6cm}
\psfig{figure=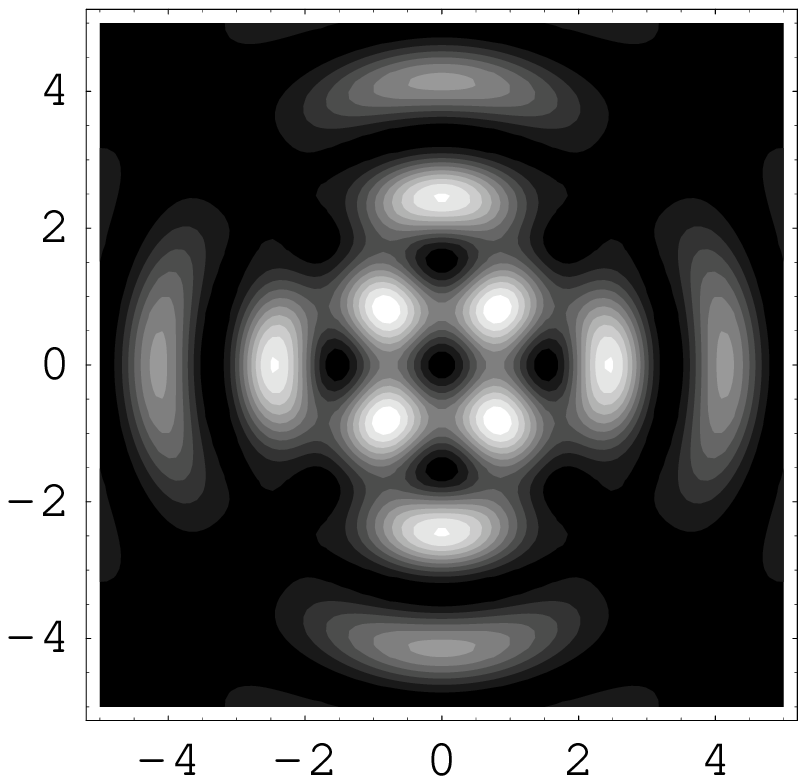,height=6cm}}
\caption{
Spatial variation of the local tunneling conductance, centered at a
non-magnetic impurity in a 2D f-wave superconductor with
$\vec{\Delta} ({\bf k}) = \Delta  \hat{z} \cos{(2 \phi)} \exp{(\pm i \phi )}$.
On the left hand side, the dominant contribution
$|u({\bf r})|^2$ at the
positive bound state resonant frequency $E_0$ is shown.
On the right hand side, the dominant contribution
$|v({\bf r})|^2$ at the
negative bound state resonant frequency $-E_0$ is shown.
The solution in this figure corresponds to the weak impurity
scattering limit.
}
\end{figure}

\begin{figure}[h]
\centerline{\psfig{figure=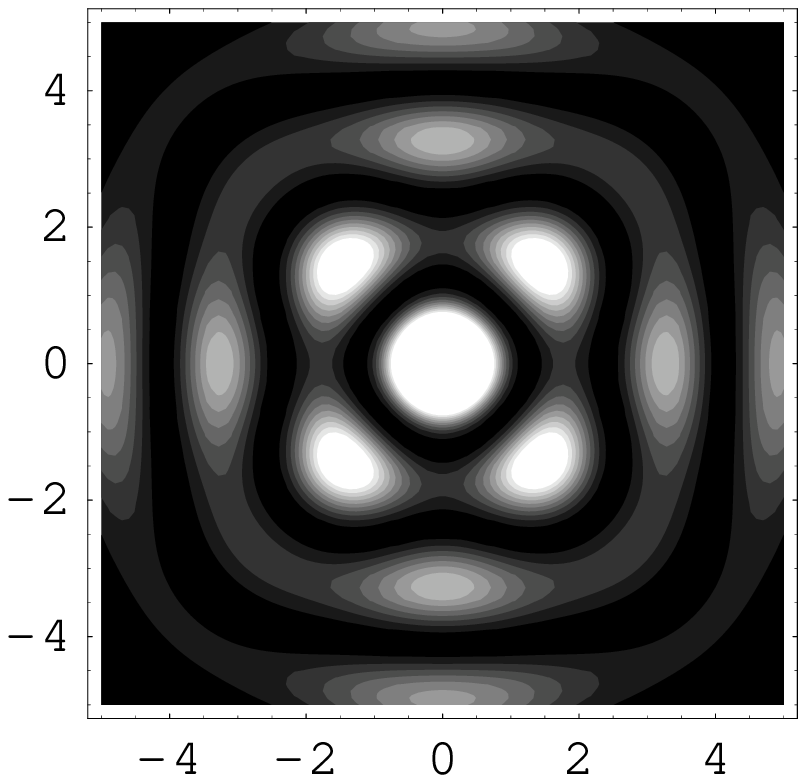,height=6cm}
\psfig{figure=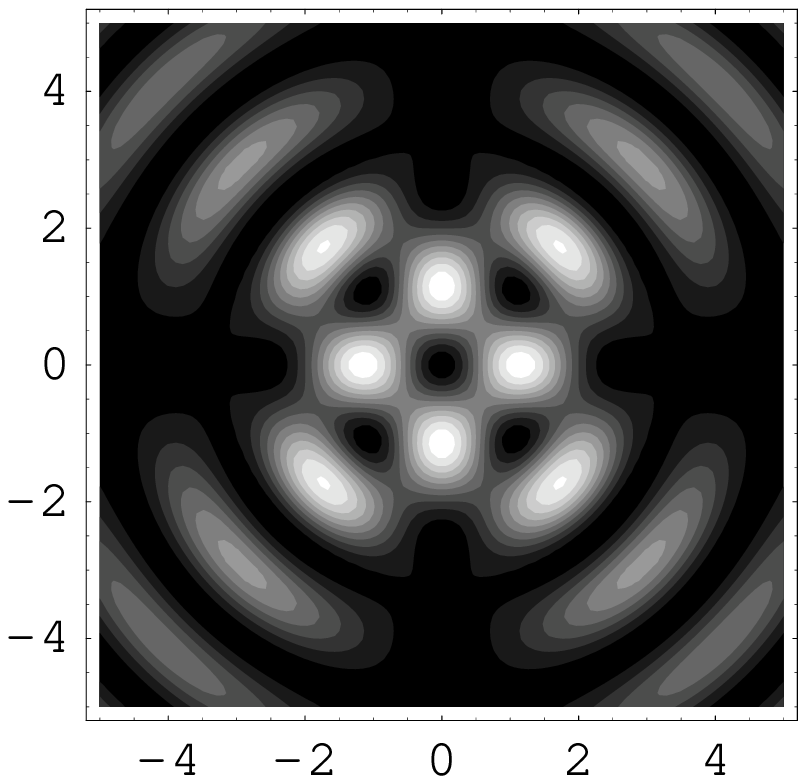,height=6cm}}
\caption{ 
Same as Fig. 2, but for strong impurity scattering.
}
\end{figure}

\end{document}